\documentstyle[psfig]{mn}

\title[blue/UV continuum shape and the ratio of radio to optical
emission]
{Relation between blue/UV continuum shape and the ratio of radio to optical emission for B3-VLA
quasars}
\author[M.F.Gu et al.]
       {Minfeng Gu, Xinwu Cao, D. R. Jiang and Y. Xu\\
1.Shanghai Astronomical Observatory, Chinese Academy of Sciences,
Shanghai, 200030, China; gumf@center.shao.ac.cn\\ 2.National
Astronomical Observatory, Chinese Academy of Sciences, Beijing,
China} \pagerange{\pageref{firstpage}--\pageref{lastpage}}
\pubyear{}

\begin{document}
\maketitle
\label{firstpage}

\begin{abstract}
The low frequency radio luminosity is believed to be an indicator
of jet power, while the optical/UV emission is probably from
accretion discs in the nuclei of steep-spectrum radio quasars. We
present a correlation between the ratio of radio to optical
luminosities and the continuum spectral index in blue/UV bands,
which might indicate that the continuum shape in blue/UV bands is
related to the ratio of jet to accretion power. The results may
imply that the spectra and structure of accretion discs are
probably affected by the interactions between jets and discs.

\end{abstract}


\section{Introduction}
The optical/ultraviolet continuum in quasars is characteristic of
big blue bump (Elvis et al. 1994; Sanders et al. 1989) and is
conventionally parameterized by a power-law of the type of
$f_{\nu}\propto\nu^{\alpha}$. Thermal emission from accretion
discs has long been the standard paradigm for the
optical/ultraviolet continuum in quasars. Many different
calculations on disc spectrum have been performed to fit observed
optical/UV continuum, which includes the simplest, bare,
geometrically thin and optically thick non-relativistic accretion
disc models (Shields 1978; Malkan 1983), the relativistic spectrum
calculations of a disc surrounding a Kerr black hole (Laor \&
Netzer 1989; Laor et al. 1990; Natali et al. 1998), and even the
calculations that both the vertical structure and radiation
transfer of the disc are considered on the assumption of the
non-local thermodynamic equilibrium (NLTE) (Hubeny \& Hubeny 1997;
1998). The observed quasar continuum can usually be fitted by
theoretical model calculations.

Radio-loud quasars are characteristic of jets, and the power of
the jet is believed to be extracted from the accretion disc/black
hole rotation energy (Blandford \& Znajek 1977; Blandford \& Payne
1982). Recently, numerical smulations show that the jet can be
accelerated from the disk region very close to the black hole
(Koide et al. 1999). Donea \& Biermann (1996) have performed
theoretical disc spectrum calculations in the presence of the jet,
though no significant difference is found in the optical/UV
continuum between radio-loud and radio-quiet quasars (Miller et
al. 1993). The optical/UV continuum of flat-spectrum quasars may
be enhanced by relativistic beaming of the jet, which prevents us
from probing the accretion discs directly through their optical/UV
continuum. An alternative approach is applying optical line
emission instead of the continuum emission as an indicator of
accretion disc for flat-spectrum quasars (Rawlings et al. 1989;
1991). The extended low frequency radio emission can be taken as
an indicator of jet power (Baum et al. 1989; Rawlings et al.
1989). The SED of the accretion discs can be studied directly from
their observed optical/UV continuum for steep-spectrum quasars
(Serjeant et al. 1998) because their optical/UV continuum is
believed to be from accretion disc without being contaminated by
the synchrotron emission from the jet.

Resently, Carballo et al. (1999) present a study on the shape of
the blue/UV continuum of quasars from the B3-VLA Quasar Sample
(Vigotti et al. 1997) selected in radio at 408 MHz. Their work is
based on $UBVR-$photometry of 73 quasars from the B3-VLA Quasar
Sample. In this paper, we study the relation between the shape of
the blue/UV continuum and the ratio of radio to optical
emission at 2400 \AA \hspace{1 mm} for Carballo's sample. In
Section 2, we briefly describe the sample. The results obtained
are included in Section 3. The last Section is the discussion of
the results. The cosmological parameters $H_{0}=50$ km s$^{-1}$
Mpc$^{-1}$ and $q_{0}=$0.5 have been adopted in this work.

\section{The  sample}
The sample of quasar candidates and the final B3-VLA Quasar Sample
are well described in Vigotti et al. (1997). Carballo et al.
(1999) performed the $UBVR$ photometric observations on a
representative group of 73 quasars from the B3-VLA Quasar Sample.
Their sample is equivalent to the B3-VLA quasar sample (except for
the RA constraint) for radio flux density $S_{408}>$0.4 Jy and
includes only a few quasars with fainter radio fluxes than the
limit, but generally close to the limit. The optical completeness
of their sample is therefore about 80 per cent (see details in
Carballo et al. 1999). Empirically, the shape of the optical/UV
continuum of quasars is generally parameterized by a power-law,
though the overall shape would be more complicated than a
power-law.  Carballo et al. (1999) use the
available photometry at the four bands, $UBVR$, to fit the
continuum of the sources by $\chi^{2}$ minimization, and the
spectral indices of 61 quasars are given in their Table 4. Those
spectral indices are all fitted with photometric error weighted
power-law model. Among all 61 spectral index derived sources, 41 have
acceptable power-law fits and the remainder have some large
$\chi^{2}$ minimum with power-law fits (private communication with
Carballo). In this work, we use their spectral index to indicate
the Blue/UV continuum shape.

For different redshifts, $UBVR$ observations represent fluxes in
different wavelength ranges in the rest-frame of sources. The flux
at some fixed wavelengths was used to construct composite spectrum
by Carballo et al. (1999). The composite spectrum is normalized by
the flux at 2400\AA \hspace{1 mm} to include the largest number of
quasars. Carballo et al. (1999) then use $L_{2400}$ to study the
relations between blue/UV luminosity and other quantities. In this
work, we use almost the same approach to derive the luminosity at
2400\AA \hspace{1 mm} in rest frame, the flux density at
2400$(1+z)$\AA \hspace{1 mm} are obtained by linear interpolation
or extrapolation between the two nearest data points (slightly
different from Carballo et al. 1999), then the luminosity at
2400\AA \hspace{1 mm} with K-correction $C=-$log$(1+z)$ is
available. The radio luminosity at 408MHz is given by K-correction
$C=-(1+\alpha)$log$(1+z)$, where $\alpha$ is the two-point radio
spectral index ($f_{\nu}\propto\nu^{\alpha}$) between 408 and
1060MHz given by Vigotti et al. (1989).

The B3-VLA Quasar Sample is selected at low frequency 408MHz,
where the steep-spectrum quasars dominates. We estimate the
two-point spectral index $\alpha_{\rm r}$
($f_{\nu}\propto\nu^{\alpha_{\rm r}}$) between 4.85GHz and 10.6GHz
with available data (Vigotti et al. 1999) and find 42 quasars with
$\alpha_{\rm r}<-0.5$ and 17 quasars with $\alpha_{\rm r}>-0.5$ in
61 spectral index-available quasars.

\section{Results}
The present sample of 61 quasars includes two strange sources with
very red spectral energy distribution (SED). We rule out them in
our statistic analysis as Carballo et al. (1999). We check the
relations between radio and optical luminosities and fluxes,
respectively and find that the correlations are still hold both in
luminosities and fluxes for the sources with $\alpha_{\rm
r}<-0.5$, which is similar to that of the whole sample (Carballo
et al. 1999).

We investigate the relation between the optical spectral index
$\alpha_{\rm o}$ and the ratio of the radio emission at 408MHz to
the optical emission at 2400\AA \hspace{1 mm}in Figure 1. A
significant anti-correlation is found at 99.4 per cent confidence
using spearman's correlation coefficient $\rho$ and a slightly
lower anti-correlation at 98.65 per cent for the sources with
$\alpha_{\rm r}<-0.5$.

A similar correlation is found between the optical spectral index
$\alpha_{\rm o}$ and the ratio of radio luminosity at 4.85GHz to
optical luminosity both for the whole sample and the subsample
with $\alpha_{\rm r}<-0.5$.

We also analyze the relation between the optical index
$\alpha_{\rm o}$ and the luminosity at 2400\AA \hspace{1 mm}in the
rest-frame and find that there is only a weak correlation (at 91.4
per cent confidence) for the whole sample, while no correlation
(at 80 per cent confidence) is present for the sources with
$\alpha_{\rm r}<-0.5$.

\section{Discussion}
The relation between the jets and the accretion processes in the
central 'engine' is a crucial ingredient in our understanding of
the physics of active galactic nuclei. Jets in quasars may be
powered by rotating black hole in the nuclei (Blandford \& Znajek
1977; Moderski et al. 1998), or be accelerated by the magnetic
field of accretion disc (Blandford \& Payne 1982). More recently,
Ghosh \& Abramowicz (1997), Livio et al. (1999) have shown that
the electromagnetic output from the inner disc is generally
expected to dominate over that from the black hole. If this is the
case for present sample, the jet power $Q_{\rm {jet}}$ is mainly
determined by the properties of the disc, say, the magnetic field
strength or/and the disc structure. Meanwhile, the shape of the
disc spectrum is also determined by the physical quantities of the
disc. It is shown that the wind from the disc would change the
temperature distribution in the inner region of the disc
significantly (Knigge 1999), which will surely affect the shape of
disc spectra.

The low frequency radio emission of quasars can be a good
indicator of the jet power $Q_{\rm {jet}}$, especially for
steep-spectrum quasars. The radio emission of steep-spectrum
quasars is believed to be mainly from the unbeamed lobes and may
well reflect the $Q_{\rm {jet}}$ (Rawlings \& Sanders 1991). For
steep-spectrum quasars, the optical continuum is believed to be
mainly from accretion disc and the continuum shape of these
sources is mainly determined by the properties of the disc, such
as, the temperature distribution and radiation transfer, etc.

The ratio of radio to optical continuum at 2400\AA \hspace{1
mm}may reflect the ratio of $Q_{\rm {jet}}$ to accretion power
$Q_{\rm {disc}}$ to some extent for steep-spectrum quasars. The
correlations found here between the optical spectral index and the
ratio of radio to optical luminosity may imply that the shape of
the disc spectrum is related to the ratio of $Q_{\rm {jet}}/Q_{\rm
{disc}}$. The sources with softer continuum (lower $\alpha_{\rm
o}$) have the higher ratio of $Q_{\rm {jet}}/Q_{\rm {disc}}$.
Knigge (1999) calculates the disc temperature distribution with
the different mass-loss winds and finds that the temperature of
the inner regions of the disc would drop dramatically for strong
winds. It may be the case for massive black holes, though his
calculations is performed for stellar black holes. The correlation
we obtain can therefore be tentatively explained by these
theoretical results. The higher ratio of radio to optical
continuum may correspond to the case of higher mass-loss rate
(Knigge 1999) which means the lower temperature in the inner
regions of the disc and then leads to the softer emitted spectrum.
The further calculations on the emitted spectrum of the disc with
winds surrounding a massive black hole is strongly desired, which
would be helpful in understanding the correlation presented here.

There are several other effects that may lead to present
anti-correlation between the optical spectral index $\alpha_{\rm
o}$ and the ratio of the radio emission to the optical emission at
2400\AA \hspace{1 mm}. Firstly, it is notable that the more
optical luminous quasars have harder spectra (O'Brien et al. 1988;
Francis et al. 1991; Carballo et al. 1999). If the radio emissions
are independent of the optical emissions,  a luminous quasar in
the optical bands tends to have a small ratio of radio to optical
emission. The lack of a significant correlation between the
optical spectral index and luminosity seems to rule out this
possibility at least for this sample. Secondly, the dust in the
sources may play some roles on the observed fluxes in the optical
bands even after the Galactic extinction correction.  More dust in
the line of sight may cause more absorption and reddening of the
spectrum in the optical bands. If the intrinsic luminosities of
the sources at 2400\AA \hspace{1 mm} in the rest-frame have no
significant difference and the radio emission is independent of
the optical spectral index, the dust reddening will lead to this
anti-correlation. It is difficult to estimate the dust effects
quantitatively, since we are not clear on the intrinsic luminosity
and the dust extinction. However, no correlation between optical
spectral index and luminosity implies that the present
anti-correlation may not be caused by dust reddening, though it
may play some roles in it. A further study basing on larger sample
of steep-spectrum quasars with photometry in multi-wavebands from
the infrared to the UV (even soft X-ray) is necessary to check
whether such an anti-correlation is still hold.

\section*{Acknowledgments}
We would like to thank the referee's comments. Ruth Carballo is
thanked for helpful discussion on the derivation of the optical
spectral index. This work is supported by the NSFC and Pandeng Project.

{}

\begin{figure}
\centerline{\psfig{figure=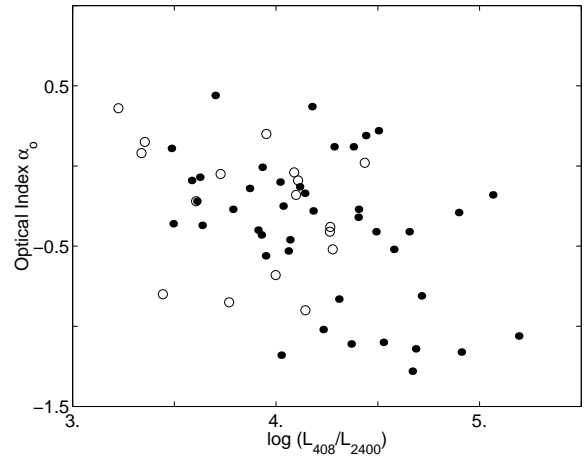,height=6.0cm}} \caption{The ratio
of radio to optical luminosity $L_{408}/L_{2400}$ and the optical
spectral index $\alpha_{\rm o}$ relation. The open circles
represent the quasars with $\alpha_{10.6-4.85}> -0.5$, and the
full circles represent the quasars with $\alpha_{10.6-4.85}\le
-0.5$ }
\end{figure}

\end{document}